\documentclass[amsmath,amssymb]{revtex4-2}
\usepackage[colorlinks=true, linkcolor=red, citecolor=blue, urlcolor=blue]{hyperref}
\usepackage[utf8]{inputenc}
\usepackage{graphicx}
\usepackage{algorithm}
\usepackage{algpseudocode}
\usepackage{amssymb}
\usepackage{amsmath}
\usepackage{bm}
\usepackage{dsfont}
\usepackage{tikz-cd}
\usepackage{enumerate}
\usepackage{stmaryrd}
\usepackage[percent]{overpic}
\usepackage{mathtools}
\usepackage{upgreek}

\newcommand{\ex}[1]{{\left\langle{#1}\right\rangle}}
\newcommand{\ket}[1]{\left|{#1}\right\rangle}
\newcommand{\bra}[1]{\left\langle{#1}\right|}

\newcommand{\qi}{\mathbf{i}}
\newcommand{\qj}{\mathbf{j}}
\newcommand{\qk}{\mathbf{k}}
\newcommand{\RN}{\mathbb{R}}
\newcommand{\CN}{\mathbb{C}}
\newcommand{\QN}{\mathbb{H}}
\newcommand{\ON}{\mathbb{O}}
\newcommand{\SN}{\mathbb{S}}

\begin{document}
\title{The Standard Model Symmetry and Qubit Entanglement}

\author{Jochen Szangolies}
\email{Jochen.Szangolies@dlr.de}
\affiliation{Institute for Software Technology, German Aerospace Center (DLR), Cologne, Germany}

\begin{abstract}

Research at the intersection of quantum gravity and quantum information theory has seen significant success in describing the emergence of spacetime and gravity from quantum states whose entanglement entropy approximately obeys an area law. In a different direction, the Kaluza-Klein proposal aims to recover gauge symmetries by means of dimensional reduction of higher-dimensional gravitational theories. Integrating both, gravitational and gauge degrees of freedom in $3+1$ dimensions may be obtained upon dimensional reduction of higher-dimensional emergent gravity. To this end, we show that entangled systems of two and three qubits can be associated with $5+1$ and $9+1$ dimensional spacetimes respectively, which are reduced to $3+1$ dimensions upon singling out a preferred complex direction. In the latter case, this reduction is invariant under a residual $SU(3) \times SU(2) \times U(1) /\mathbb{Z}_6$ symmetry, the Standard Model gauge group. This motivates a picture in which spacetime emerges from the area law-contribution to the entanglement entropy, while gauge and matter degrees of freedom are due to area law-violating terms. We remark on a possible natural origin of the chirality of the weak force in the given construction. Furthermore, we highlight the possibility of using this construction in quantum simulations of Standard Model fields.

\end{abstract}

\maketitle
\section{Introduction: The `Quantum First'-Program}

If our world, as is widely believed, is quantum, it is in principle representable as a single vector evolving in Hilbert space, and thus, all spatiotemporal notions must be reducible to this description. This suggests a research program that has been variously called `Hilbert Space Fundamentalism' \cite{carroll2021reality}, `Mad Dog Everettianism' \cite{carroll2019mad}, `extreme Occam's razor' \cite{tegmark2015consciousness}, or simply `physics from scratch' \cite{tegmark2008mathematical}. Essentially, rather than trying to find quantized versions of classical notions, whether of spacetime or of classical fields, one instead starts with a quantum description, and aims to locate the appropriate dynamics within. We will thus think of it as simply the `quantum first'-program (cf. \cite{giddings2019quantum}).

This is a formidable task. Hilbert space itself is exhaustively described by its dimension; a vector in Hilbert space contains virtually no information regarding its physical content---a single-qubit space $\mathbb{C}^2$ might model an electron's spin just as well as a current direction in a Josephson junction; a vector $\ket{\psi}$ in this space might designate spin up, down, or any superposition of these.

Yet, despite this apparent paucity of information, important advances have been made. Most significantly, a thriving research program has established surprising links between quantum mechanics and general relativity, outlining the possibility of deriving space and time, as well as its dynamics, from an abstract quantum description. Maldacena and Susskind have proposed the $ER=EPR$ conjecture \cite{maldacena2013cool}, conjecturing that quantum entanglement is equivalent to wormholes in spacetime. (However, it should be noted that due to the quantum no-communication theorem \cite{eberhard1989quantum}, these wormholes are necessarily non-traversable, thus preserving the locality underlying general relativity). Indeed, the links appear so deep that Susskind has proposed to generalize $ER = EPR$ to $GR=QM$, stating that `where there is quantum mechanics, there is also gravity' \cite{susskind2017dear}. Hence, a quantum system does not merely come equipped with a geometry, but that geometry is dynamical, approximating general relativistic spacetimes in a classical limit.

The origin of this idea can be traced to work of Jacobson, who has shown that Einstein's equations can be recovered from thermodynamics in the presence of black hole horizons \cite{jacobson1995thermodynamics} (see also \cite{padmanabhan2010thermodynamical,verlinde2011origin}). These arguments rely essentially on the area law-scaling of the Bekenstein-Hawking entropy of black holes \cite{bekenstein1973black}. This scaling is also observed, in many cases, for the entanglement entropy of quantum systems---indeed, it has been proposed to explain black hole entropy as the entanglement entropy of degrees of freedom across the horizon \cite{bombelli1986quantum,srednicki1993entropy}.

Along a different thread, van Raamsdonk has suggested that entanglement may be responsible for spacetime connectivity \cite{van2010building,vanraamsdonk2018building}. In the AdS/CFT correspondence, the Ryu-Takayanagi formula \cite{ryu2006aspects} relates the entanglement between different patches of the conformal field theory on the boundary of Anti-de Sitter space with the area of a surface partitioning the AdS-space; if the entanglement between both goes to zero, the space partitions into two disconnected regions. 

Combining these perspectives, the linearized Einstein equations follow from `entanglement thermodynamics' for perturbations of the AdS-spacetime \cite{lashkari2014gravitational,faulkner2014gravitation}. While the bulk of this work assumes the AdS/CFT-correspondence, Jacobson has shown that a derivation of the full Einstein equations is possible without assuming an AdS-background, given a condition of entanglement equilibrium \cite{jacobson2016entanglement}. 

A way to think about this is in terms of a graph, each node of which is itself an abstract element of quantum information---a qubit, or perhaps more generally, some finite-dimensional Hilbert space---that carries no spatial information itself, with the edges encoding the entanglement structure of the whole system. Geometry, then, must be reconstructed from this basic picture.

This is, essentially, the proposal made by Cao and Carroll \cite{cao2018bulk} (see also \cite{cao2017space}). Starting from a graph as described above, they show that the metric structure of spacetime, as well as the linearized Einstein equations, can be obtained, using an argument closely related to that of Jacobson.

\begin{figure}[h] 
 \centering
 \begin{overpic}[width=0.5\textwidth]{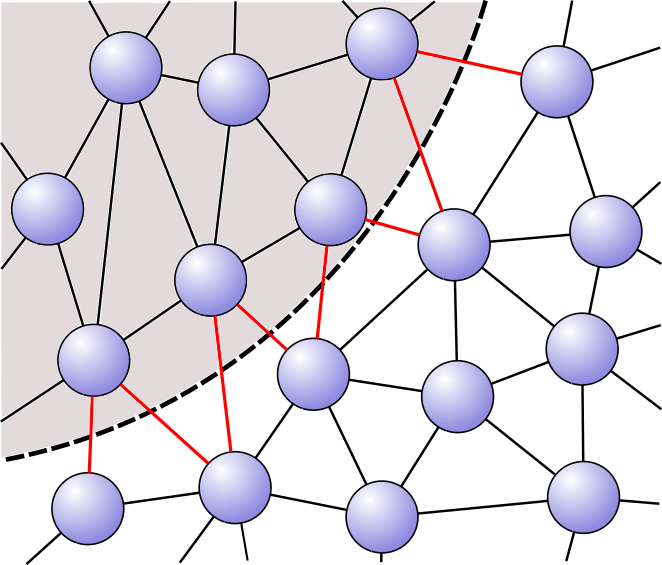}
  \put(3,81){$R$}
  \put(95,81){$\bar R$}
  \put(74,81){$\partial R$}
 \end{overpic}
\caption{The entanglement entropy of a region $R$ across its boundary $\partial R$ is computed by summing up the contributions of entanglement `links' cut by the boundary. For a state with `short-range' entanglement, this yields an area law.}
\label{pic:RC}
\end{figure}

Intriguingly, there appears a natural bipartitioning of the entanglement degrees of freedom in proposals such as the one in Ref.~\cite{cao2018bulk}. Recall that the area-law scaling of entanglement entropy plays a crucial role in recovering spacetime. Such a scaling is typically only observed in a very special class of states, such as the ground states of local gapped many-body systems \cite{eisert2010colloquium}. In Ref.~\cite{cao2018bulk}, a generalized notion termed `redundancy-constrained' (RC) states is used, which correspond to graphs such that the entanglement entropy can be computed by summing over all the contributions of nearest-neighbor links over a given bipartition of the system into a region $R$ and its complement, $\bar{R}$. Explicitly,
\begin{equation}
    S_{RC}(R) = \frac{1}{2}\sum_{i\in R, j\in \bar{R}}I(i:j),
\end{equation}
where $I(i:j)$ is the quantum mutual information between nodes $i$ and $j$ (see Fig.~\ref{pic:RC}).

One can now consider states that only approximately satisfy this constraint. Then, the entanglement entropy receives a subleading correction
\begin{equation}
    S(R) = S_{RC}(R) + S_{sub}(R),
\end{equation}
which can be associated with long-range entanglement or excited degrees of freedom over and above the vacuum. This subleading correction may then be associated to the entanglement entropy of an emergent gauge theory \cite{lin2018ryutakayanagi}.

Along similar lines, Cao \cite{cao2021quantum}, following work by Harlow \cite{harlow2017ryu}, has proposed to interpret the splitting of contributions to the entanglement entropy in terms of quantum error correcting codes (QECC)---specifically, QECC protecting against subsystem erasure. There, the information in some state $\rho$ is nonlocally encoded into $\Tilde{\rho}$ in such a way as to be protected against erasure of the subsystem $\bar{R}$. Thus, there exists a decoding unitary on $R$
\begin{equation}
    U_R\Tilde{\rho} U_R^\dagger = \rho\otimes \chi,
\end{equation}
where $\chi$ is the generically entangled state of the remaining degrees of freedom, that restores the encoded information $\rho$. Thus, the full information encoded on $R\bigcup\bar{R}$ can be recovered from $R$ alone. This yields for the entropy of $\Tilde{\rho}_R = \mathrm{Tr}_{\bar{R}}(\Tilde{\rho})$:
\begin{equation}
    S(\Tilde{\rho}_R) = S(\chi) + S(\rho),
\end{equation}
where $S(\chi)$ can be interpreted as the geometric, area-law contribution to the entropy, and $S(\rho)$ in turn as stemming from the emergent matter fields.

\begin{figure}[h] 
 \centering
 \begin{overpic}[width=0.5\textwidth]{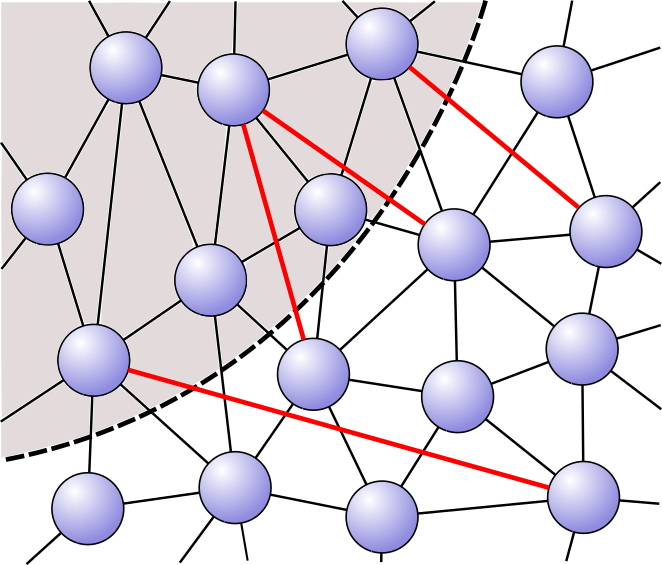}
  \put(3,81){$R$}
  \put(95,81){$\bar R$}
  \put(74,81){$\partial R$}
 \end{overpic}
\caption{A state with additional long-range entanglement, or entanglement due to a state `encoded' into it in the sense of quantum error correction, receives subleading corrections to the entanglement entropy from these additional terms.}
\label{pic:RCapprox}
\end{figure}

It is here that our investigation begins. If dynamical spacetime emerges from the area-law contribution to the entanglement, it seems a natural proposal to investigate the emergence of matter degrees of freedom, together with their symmetries, from the entanglement yielding subleading corrections to the entropy---the `logical', encoded state $\rho$, on the above view. However, we stress that this motivation is, at this point, heuristic; the discussion in the sequel is independent of this interpretation.

Our main result is then to show how to use a construction of Dubois-Violette and Todorov \cite{dubois2016exceptional,dubois2019exceptional}, as extended and clarified by Krasnov \cite{krasnov2020so9}, to derive the exact symmetry and degrees of freedom of (a generation of) Standard Model fermions from the entanglement properties of a three-qubit state. This proceeds by two routes: first, the internal symmetries of a left-handed half-generation are derived from the transformations leaving a preferred complex structure invariant. Then, via a kind of `dimensional reduction', the three-qubit state yields the spacetime-degrees of freedom of a fermion generation, together with the internal symmetries of the right-handed part.

One can think of this in the following way. Standard, complex quantum mechanics can, as noticed long ago by von Weizsäcker \cite{weizsacker1958komplementaritat} (whose `ur-theory' \cite{von2006structure} might be considered the first example of a `quantum first'-approach), be associated with $3+1$-dimensional spacetime. In more modern times, it has been shown that the three-dimensionality of position space and the state-space of elementary quantum systems, and the necessity of their correspondence, follow from information-theoretical considerations \cite{mueller2013three,dakic2016classical}, and indeed, even the Lorentz transformations can be obtained by considering the relation between different experimenters' descriptions \cite{hohn2016operational}.

These arguments can be generalized to quantum theory over the quaternions $\QN$, leading to a $5+1$-dimensional spacetime. Furthermore, extending to the octonions $\ON$, we obtain $9+1$ dimensions \cite{brody2011six} (which are, not coincidentally, precisely the dimensions in which a supersymmetric Lagrangian exists \cite{baez2009division,baez2011division}). 

These spacetimes can then be broken down to $3+1$ dimensions by simply singling out a preferred complex structure---that is, fixing a split of $\QN$ into $\CN\oplus\CN$ and $\ON$ into $\CN\oplus\CN^3$. It is then the observation of Dubois-Violette, Todorov, and Krasnov \cite{dubois2016exceptional,dubois2019exceptional,krasnov2020so9} that these splits are left invariant by $SU(2)\times U(1)/\mathbb{Z}_2$ for the quaternionic case and $SU(3)\times SU(2)\times U(1)/\mathbb{Z}_6$ for the octonionic one. This can be compared to a `quantum first' realization of the proposal due to Kaluza \cite{kaluza1921unitatsproblem} and Klein \cite{klein1926atomicity,klein1926quantentheorie} that higher-dimensional spacetime theories can lead to gauge fields in lower dimension, with the origin of the gauge symmetry being a $Spin(5)$- respectively $Spin(9)$-action on the two- and three-qubit state space. 

The appeal to higher division-algebraic generalizations of quantum theory has a long and distinguished history. Of note, Finkelstein, Jauch, Schiminovich and Speiser provided the first example of quaternionic quantum theory in 1962/1963 \cite{finkelstein1962foundations,finkelstein1963principle}---the latter of which is remarkable for introducing a mass-generating mechanism for an $SU(2)$-gauge theory by means of a quaternion-imaginary dynamical field playing much the same role as the Higgs field in the later explanation of electroweak symmetry breaking. 

Applications of the octonions to particle physics date back to work by G\"unaydin and G\"ursey, who first proposed using them to model the strong force \cite{gunaydin1974quark}. The attractivity of appealing to the (normed) division algebras $\CN $, $\QN $ and $\ON $ comes from the observation that the unit complex numbers form the group $U(1)$, the electromagnetic gauge group, while the unit quaternions form $SU(2)$, the gauge group of the weak interaction. Of course, due to their non-associative nature, the unit octonions form no group at all---but still, their automorphism group is the exceptional $G_2$, which contains $SU(3)$ (the gauge group of the strong interaction) as a subgroup---in fact, the subgroup preserving one octonion imaginary direction, and thus, singling out a preferred complex structure. 

This has led to an alternative, algebraic approach to models of particle physics. Dixon, for instance, focuses on the algebra $\mathbb{D}=\RN \otimes\CN \otimes\QN \otimes\ON $ (where the factor of $\RN $ is technically redundant and only included for aesthetic purposes), and uses it to build a model incorporating the complete matter and force content of the Standard Model \cite{dixon2013division}. More recently, Furey \cite{furey2012unified,furey2018three} has proposed her unified theory of ideals, drawing on actions of the Dixon algebra $\mathbb{D}$ on itself. Stoica \cite{stoica2018leptons} has proposed a construction in which the Standard Model emerges from the complex Clifford algebra $\mathbb{C}\ell_6$, with the color- and electromagnetic symmetry appearing in the same way as in the model of Furey. Also of note here is the model of Morita \cite{morita1982quaternionic,morita1981octonions,morita1982algebraic}, giving a quaternionic version of the theory of electroweak interactions, and an octonionic QCD. For a more thorough review of this `algebraic' approach to the Standard Model, see \cite{stoica2018leptons}.

The final piece of the puzzle then is the origin of quaternionic/octonionic quantum theory. Here, we appeal to work by Mosseri and Dandoloff \cite{mosseri2001geometry,mosseri2006two} and Bernevig and Chen \cite{bernevig2003geometry}: entangled states of two qubits can be written in terms of quaternionic spinors---more accurately, the state space of a two-qubit system is given by the quaternionic Hopf fibration (just as the state space of a single qubit, the Bloch sphere, is the complex Hopf fibration). Likewise, three-qubit systems can be considered as octonionic spinors, whose state space is given by the octonionic Hopf fibration. This also provides the point of origin for the emergence of the Standard Model gauge group: the equivariance group of the octonionic Hopf fibration is $Spin(9)$, whose subgroup respecting the splitting $\ON\simeq\CN\oplus\CN^3$ is just $SU(3)\times SU(2)\times U(1)/\mathbb{Z}_6$ \cite{krasnov2020so9}.

Furthermore, the Hopf maps in each case are entanglement sensitive---disentangling each qubit results in the space decomposing into a product of individual qubit state spaces. The information within separable states is then fully representable within ordinary, empty $3+1$-dimensional spacetime; but entanglement introduces additional, `internal' degrees of freedom. The heuristic picture we propose then is one where $3+1$ dimensional spacetime with its dynamics emerges from the area-law contribution to the entanglement of a quantum state, while area-law violating corrections can be viewed as `bubbles' of higher-dimensional spacetime, which, upon reduction to $3+1$ dimensions, which can be interpreted as singling out a preferred tensor product decomposition, add matter and gauge degrees of freedom. 

The origin of the Standard Model symmetry and its attendant idiosyncrasies has been an active topic of research since its inception. Grand Unified Theories (GUTs) have proposed to embed its gauge group into a larger one, recovering its phenomenology after appropriate symmetry breaking; however, there is a large number of such schemes conceivable, with little guidance for choosing among them. The division algebraic approach reduces this embarrassment of riches somewhat, but even here, finding a unique model has been challenging, and the reason for their appearance---what, if anything, singles them out as relevant to the description of elementary particles---is not well understood. 

Our main aim here is to provide a possible physical picture that offers some guidance. In particular, as we will see, the division algebraic description is singled out by the geometry of qubit Hilbert spaces, potentially significantly reducing the search space for theory-building. Furthermore, by building a bridge between the `quantum first' and division algebraic programs, we hope that the insights of each may help stimulate research within the other, in particular by introducing the possibility of emergent matter (and gauge) degrees of freedom, but more generally by providing a unified framework for both. 

The remainder of this article is structured as follows. In Sec.~\ref{sec:compl}, we introduce the basic template the subsequent three sections will follow. First, we give a brief introduction to the scenario, starting with a single qubit; then, in Sec.~\ref{sub:complspace}, we present the connection of complex quantum mechanics with $3+1$-dimensional spacetime. In the following Sec.~\ref{sub:comphopf}, we introduce the complex Hopf fibration, and show its relation to the state space of single qubits. Then, in Sec.~\ref{sub:compspin}, we illustrate the importance of the group $Spin(3)$ as the complex Hopf map's equivariance group. 

In Sec.~\ref{sec:quat}, the construction is generalized to two qubits, and quaternions, showing the connection to $5+1$-dimensional spacetime (\ref{sub:quaspace}), and the quaternionic Hopf fibration (\ref{sub:quahopf}). In Sec.~\ref{sub:quaspin}, then, it is illustrated how the group $SU(2)\times U(1)/\mathbb{Z}_2$ emerges from the subgroup of $Spin(5)$ leaving a quaternion imaginary direction invariant. In Sec.~\ref{sec:oct}, the construction is analogously generalized to the octonions and three qubits, $9+1$-dimensional spacetime (\ref{sub:octspace}), the octonionic Hopf fibration and its relevance for three-qubit entanglement (\ref{sub:octhopf}), and finally, the emergence of $SU(3)\times SU(2) \times U(1)/\mathbb{Z}_6$ as leaving a choice of complex structure invariant (\ref{sub:octspin}).

In Sec.~\ref{sec:mass}, we will then speculate about an extension of the model to incorporate a full generation of Standard Model fermions by generalizing to the four-qubit case, which will also allow us to include massive states in our model, and yield a possible natural explanation for the chirality of the weak force. As the presented construction develops a correspondence between gauge fields and few-qubit Hilbert spaces, Sec.~\ref{sec:Simul} discusses the possibility of using it for simulations on current quantum computers. Finally, we conclude in Sec.~\ref{sec:conc}, and propose some directions of further development, such as the incorporation of a Higgs mechanism by taking account of the non-triviality of the embedding of the fibre in the Hopf fibration.

\subsection{Outline of the Approach}

The following Sections~\ref{sec:compl}, \ref{sec:quat}, and \ref{sec:oct} all follow a similar template, which we outline here for clarity. After recapitulating some elementary details on the mathematical description of the systems under consideration (single, respectively two- and three-qubit systems), we rehearse an argument that links their state-spaces to the symmetries of Minkowski spacetime of varying spatial dimension. This takes its leave from what von Weizsäcker has called `abstract quantum theory' \cite{drieschner1988reconstruction}: that is, individual qubits are not considered as spatiotemporally located physical systems (e.g., electron spins), but rather, their purely abstract properties as linked to their mathematical description are the subject of examination. (Von Weizsäcker considered these as `elementary questions' that are subject to empirical investigation.) The symmetries of Minkowski spacetime (i.e. the $D+1$-dimensional Lorentz group) then is constructed as automorphisms of the state space in each case.

Then, each system is described in terms of the Hopf fibration. In the single qubit case, this just yields the familiar Bloch sphere picture: the two complex parameters of a single qubit yield four real numbers whose squares sum to one, thus parametrizing a point on the sphere in four dimensions, $S^3$. Since each qubit has an arbitrary phase, we can `neglect' it to yield a point on the Bloch sphere $S^2$ (which is the space whose automorphisms yield the $3+1$-dimensional Lorentz transformations). This realizes $S^3$ as $S^2$ with an $S^1$ fiber coming from the phase, which is the complex Hopf fibration. An analogous construction reduces the state space of a two- and three-qubit system to an $S^4$, respectively $S^8$ base space.

The interesting point here is that these two latter mappings are both entanglement sensitive: the additional dimensions of the base space characterize the entanglement between one qubit and the rest of the system; disentangling the states thus reduces either base to $S^2$, the Bloch sphere of the single qubit. We can thus introduce a split in the degrees of freedom of the base, between those belonging to a single qubit, and those characterizing the entanglement to the other qubit(s). This motivates then looking at the transformations leaving this split invariant, i.e. those that do not `mix' the entanglement- and qubit-degrees of freedom. These turn out to be $SU(2)\times U(1)/\mathbb{Z}_2$ for two qubits, and $SU(3)\times SU(2)\times U(1)/\mathbb{Z}_6$ for three qubits, with in each case the $SU(2)$ acting on the single qubit in the base, and the remaining transformations on the degrees of freedom characterizing the entanglement.

The question remains how to relate this mathematical fact to physical reality. Taking the above construction as it is, we obtain exclusively the \emph{internal} symmetry of the Standard Model in the three-qubit case, and indeed, any talk of spacetime symmetries could be bracketed entirely. While this is interesting, it remains to be seen how to reconnect to a spacetime description of a physical particle. This project is begun in Sec.~\ref{sec:mass}. There, we argue that there is an alternative perspective on this construction, which amounts to taking the degrees of freedom of the base qubit as being acted upon by the spacetime symmetries instead. In the case of the two-qubit state, we argue that this yields the spacetime degrees of freedom of a left- and a right-handed two-spinor, together with a $U(1)$-gauge symmetry on the right-handed part. This yields a Kaluza-Klein like construction, where spatial dimensions in a higher-dimensional space yield internal symmetries in the lower dimension. 

Thus, each instance of this construction can be either interpreted as giving the internal symmetries of a left half-generation of fermions, or as the spacetime symmetries of both a right- and left-handed fermion, together with the internal symmetries of the right-handed part, lacking the $SU(2)$-factor. Putting both together---which amounts, as the number of degrees of freedom is concerned, to adding another qubit---then provides all the right ingredients for the description of a single fermion generation.

\section{The Complex Case: Single Qubits}\label{sec:compl}

We start by treating the simplest case, that of single qubits in ordinary, complex quantum mechanics. Any pure qubit state can be written as
\begin{equation}
    \ket{\psi} = c_0\ket{0} + c_1\ket{1},
\end{equation}
where $c_0,c_1\in\CN $, $|c_0|^2+|c_1|^2=1$, and $\left\{\ket{0},\ket{1}\right\}$ denotes a standard basis of the two-qubit Hilbert space $\CN ^2$.

It will be useful to represent qubit states in terms of the Pauli matrices
\begin{align*}
    \sigma_1 & = \begin{pmatrix} 0 & 1 \\ 1 & 0 \end{pmatrix}\\
    \sigma_2 & = \begin{pmatrix} 0 & -i \\ i & 0 \end{pmatrix}\\
    \sigma_3 & = \begin{pmatrix} 1 & 0 \\ 0 & -1 \end{pmatrix},
\end{align*}
yielding
\begin{align}\label{eq:bloch}
    \nonumber\rho_\psi &= \ket{\psi}\!\!\bra{\psi} = \begin{pmatrix} c_0c_0^* & c_1c_0^* \\ c_0c_1^* & c_1c_1^* \end{pmatrix}\\
            &= \frac{1}{2}\begin{pmatrix} 1 + x_3 & x_1 - ix_2 \\ x_1 + ix_2 & 1-x_3 \end{pmatrix}\\
    \nonumber &= \frac{1}{2}\left(\mathds{1} + \mathbf{r}_\psi\cdot\boldsymbol{\sigma}\right),
\end{align}
with $x_1 = c_1c_0^* + c_0c_1^*$, $x_2 = i(c_0c_1^* - c_1c_0^*)$, and $x_3 = |c_0|^2 - |c_1|^2$. This yields the parametrization of the Bloch sphere and introduces the Bloch vector $\mathbf{r}_\psi = (x_1,x_2,x_3)$, whose components are given by the expectation values of the Pauli matrices for the given quantum state, $x_i = \left\langle\sigma_i\right\rangle_\psi$.

\subsection{$SL(2,\CN )$ and $3+1$-Dimensional Spacetime}\label{sub:complspace}

The correspondence between elementary quantum systems and $3+1$ dimensional spacetime may be explicated in various ways. Von Weizsäcker proposed a symmetry-based argument in the `50s \cite{weizsacker1958komplementaritat}, which has been echoed by Brukner and Zeilinger in more recent times \cite{brukner2003information}. M\"uller and Masanes give a more operationalistic argument, showing that probabilistically interacting systems exchanging `minimal amounts of direction information' yield a three-dimensional laboratory space. H\"ohn and M\"uller then show that the descriptions of two experimenters of such a situation will be related by an element of $O^+(3,1)$, the orthochronous Lorentz group \cite{hohn2016operational}. A similar result is obtained by Daki\'c and Brukner \cite{dakic2016classical}.

We propose a slightly different argument here, relying on the description of points on the heavenly sphere in terms of spinors going back to the two-spinor formalism of Penrose and Rindler \cite{penrose1984spinorsI,penrose1984spinorsII}. 

Consider a Lorentzian four-vector $(x_0,x_1,x_2,x_3)^T$. It lies on the light cone if $x_0^2-x_1^2-x_2^2-x_3^2 = 0$. The light cone can be equivalently represented in a spinorial way. Thus, consider Hermitian matrices of the form
\begin{align}
    \nonumber X &= \begin{pmatrix} c_0 \\  c_1  \end{pmatrix} \begin{pmatrix} c_0^* & c_1^*  \end{pmatrix} \\
    &= \frac{1}{2}\begin{pmatrix} x_0 + x_3 & x_1-ix_2 \\ x_1+ix_2 & x_0-x_3 \end{pmatrix},
\end{align}
where $c_0, c_1\in\CN$, yielding 
\begin{align*}
    x_0 &= |c_0|^2 + |c_1|^2, \\
    x_1 &= c_0^*c_1 + c_1^*c_0,\\
    x_2 &= -i\left(c_0^*c_1 - c_1^*c_0\right),\\
    x_3 &= |c_0|^2 - |c_1|^2.
\end{align*}

As $\mathrm{det}(X) = 0 \rightarrow  x_0^2 - x_1^2 -x_2^2 - x_3^2 = 0$, these parametrize the light cone. 

We note, for later, that matrices of this form always have a well-defined determinant
\begin{equation}\label{eq:divdet}
    \mathrm{det}\begin{pmatrix} a + b & d^* \\ d & a-b \end{pmatrix} = a^2-b^2-|d|^2
\end{equation}
for $a$, $b \in \RN$, and $d\in\RN,\CN,\QN$ or $\ON$ \cite{baez2002octonions}.

Taking the three-dimensional slice at $x_0 = 1$, we obtain the representation in Eq.~\eqref{eq:bloch}, and this condition is satisfied if $x_1^2 + x_2^2 + x_3^2 = 1$, which is the normalization condition for a pure-state qubit. This entails that (pure) qubits parametrize directions on the Riemann (`celestial') sphere---in other words, each pure qubit represents a light ray. For now, we do not attach any explicit physical interpretation to the 4-vector associated to each qubit, and consider them merely directions in Minkowski space; later on, it will become appropriate to think of them as giving 4-momenta of particles.

The group of automorphisms of this sphere is the M\"obius group $MG$, which transforms a complex number $z$ as follows (where $a,b,c,d$ are complex numbers such that $ad-bc\neq0$):
\begin{equation} 
     m(z) = \frac{az + b}{cz + d}.
\end{equation}
Then, the following mapping defines a homomorphism between $SL(2,\CN)$, the group of complex $2\times 2$-matrices with unit determinant, and the M\"obius group:
\begin{equation}
    \begin{matrix*}[l]
    \phi: SL(2,\CN) \longrightarrow MG \\
    M=\begin{pmatrix}
        a & b \\ c & d
    \end{pmatrix} \longrightarrow m(z) = \frac{az + b}{cz + d}
    \end{matrix*}
\end{equation}

Now consider a transformation of the matrix $X$ of the form
\begin{equation}
    X^\prime = MXM^\dagger.
\end{equation}
If $\mathrm{det}(M) = 1$, it follows that $\mathrm{det}(X^\prime) = \mathrm{det}(X)$, i. e. the transformation preserves the Minkowski interval, and is thus a Lorentz transformation. This mapping has a nontrivial kernel $\{\mathds{1},-\mathds{1}\}$, and hence, $SL(2,\CN)$ is the double cover of $SO^+(3,1)$. In this way, the symmetry of $3+1$-dimensional Minkowski space emerges from the identification of the Riemann sphere with the Bloch sphere of a single qubit.

\subsection{The Bloch Sphere and the Complex Hopf Fibration}\label{sub:comphopf}

Above, we have already introduced the Bloch sphere representation of single qubits. The normalization condition $|c_0|^2 +|c_1|^2 = 1$ implies that the Hilbert space of a single qubit is the $3$-sphere $S^3$. However, quantum states are properly represented by \emph{rays} in Hilbert space, and hence, only unique up to an overall $U(1)$ phase factor $\exp{(i\phi)}$. Taking account of this, each qubit can be represented in the projective Hilbert space $S^2 = S^3/U(1) = S^3/S^1$. Hence, there must be a map taking elements of $S^3$ to $S^2$. This is the complex Hopf map, and can be explicitly written as the composition of a map $h_1$ taking points of $S^3$ defined by complex coordinates $(c_0,c_1)$ to points on the extended complex plane $\CN \cup\{\infty\}$,
\begin{equation}
    h_1^c: \begin{matrix*}[c] S^3 & \longrightarrow & \CN \cup\{\infty\} \\ (c_0,c_1) & \longrightarrow & c_0c_1^{-1} \end{matrix*},
\end{equation}
and a map $h_2$ corresponding to an inverse stereographic projection from $\CN $ to $S^2$
\begin{equation}
    h_2^c: \begin{matrix*}[c] \CN \cup\{\infty\} & \longrightarrow & S^2 \\ c_0c_1^{-1} & \longrightarrow & (x_1,x_2,x_3) \end{matrix*},
\end{equation}
such that $x_1^2 + x_2^2 + x_3^2 = 1$. As discussed above, this is of course just the map taking a quantum state $\ket{\psi} = c_0\ket{0} + c_1\ket{1}$ to its Bloch vector representation, yielding for each coordinate \cite{bernevig2003geometry}
\begin{align*}
    x_i &= \left\langle\sigma_i \right\rangle_\psi \\
        &= (c_0^*,c_1^*)\sigma_i\begin{pmatrix} c_0 \\ c_1 \end{pmatrix},
\end{align*}
where it is obvious that each $x_i$ is defined only up to an $U(1)$ ambiguity. 

This provides a realization of the single-qubit Hilbert space in terms of the complex Hopf fibration 
\begin{equation}
    S^3 \stackrel{S^1}{\longrightarrow} S^2,
\end{equation}
which is the simplest example of a nontrivial fibre bundle, meaning that $S^3\neq S^2\times S^1$. Physically, this nontriviality implies the impossibility of assigning a consistent phase to every point on the Bloch sphere.

It is at this point that the necessity of the division algebra property enters into the picture: without it, there is no way to define $c_0c_1^{-1}$, and thus, to formulate the Hopf map unambiguously. This explains why there are precisely four such constructions, corresponding to the four division algebras $\RN$, $\CN$, $\QN$ and $\ON$. Consequently, the construction presented here has a natural endpoint at the octonionic level.

\subsection{$Spin(3)$ and the Complex Hopf Fibration}\label{sub:compspin}

We will now connect the story so far with the formalism in Ref.~\cite{krasnov2020so9}. Thus, consider matrices of the form
\begin{equation}\label{eq:spinmat}
    X(r,c)=\begin{pmatrix} r & L_c \\L_{c^*} & -r \end{pmatrix}, r\in\RN ,c\in\CN ,
\end{equation}
where by $L_c$ we mean left-multiplication by $c$ (for the commutative complex numbers, this is obviously not a salient distinction, but it will become important in the generalization to higher division algebras). This is just the matrix $\mathbf{r}_\psi\cdot\boldsymbol{\sigma}$ from Eq.~\eqref{eq:bloch} with $r = x_3$ and $c = x_1 - ix_2$.

These matrices generate the Clifford algebra $\mathbb{C}\ell_3$, as they fulfill the relation
\begin{equation}\label{eq:cl3}
    X(r,c)X(r,c) = (r^2 + |c|^2)\mathds{1}.
\end{equation}
Furthermore, these matrices generate the group $Spin(3)$ in the case of $r^2 + |c|^2 = 1$: a rotation along an axis in $\RN ^3$ can be realized by an even number of reflections, and conjugating a matrix of the form given in Eq.~\eqref{eq:cl3} by another, yielding
\begin{equation}
    X(r,c) \longrightarrow -X(r^\prime,c^\prime)X(r,c)X(r^\prime,c^\prime),\,\, r^{\prime 2} + |c|^{\prime 2} = 1,
\end{equation}
produces a reflection along ($r^\prime,c^\prime) \in \RN ^3$ \cite{krasnov2020so9}.

The group $Spin(3)$ here can also be understood as the equivariance group of the complex Hopf fibration, that is, the Hopf map is equivariant with respect to the action of $Spin(3)$. This becomes explicit by noting that the map can be written in terms of coset spaces \cite{hatsuda2009coset}:

\begin{equation}
 S^3\stackrel{S^1}{\longrightarrow}S^2 \equiv \frac{Spin(3)}{Spin(1)}\stackrel{\frac{Spin(2)}{Spin(1)}}{\longrightarrow}\frac{Spin(3)}{Spin(2)}.
\end{equation}

Here, the action of $Spin(3)$ on $S^2$ is induced by its action on $\RN ^3$, with $S^2$ just being the set of unit-length vectors of $\RN ^3$, and the action of $Spin(3)$ on $S^3$ is given by the isomorphism $Spin(3)\simeq Sp(1)\simeq U(1,\QN )$, which acts by left multiplication on (unit) quaternions, i. e. $S^3$.

\section{The Quaternionic Case: Two Qubits}\label{sec:quat}

The above construction admits a ready generalization to the quaternionic case. The skew-field of quaternions $\QN $ can be obtained from the complex numbers by means of the Cayley-Dickson construction: starting with two complex numbers $c_0 = r_0 + r_1\qi$ and $c_1 = r_2 +  r_3\qi$ (where we use bolding to mark quaternion imaginary directions), one introduces $\qj$ such that $\qj^2 = -1$, which anticommutes with $\qi$. Then, a quaternion $q_0$ is given by
\begin{align*}
    q_0 &= c_0 + c_1\qj \\
        &= r_0 + r_1\qi + r_2\qj + r_3\qk,
\end{align*}
where we have introduced $\mathbf{ij} = -\mathbf{ji} = \qk$ as the third quaternion imaginary unit. This directly entails Hamilton's relations
\begin{equation*}
    \qi^2 = \qj^2 = \qk^2 = \qi\qj\qk = -1,
\end{equation*}
which fully define quaternionic multiplication. 

Quaternionic conjugation is the operation taking $q_0$ to $q_0^*$ with
\begin{equation*}
    q_0^* = r_0 -  r_1\qi -  r_2\qj - r_3\qk,
\end{equation*}
which can be used to define the norm $|q_0|^2 = q_0q_0^*$. This allows us to define a multiplicative inverse $q_0^{-1} = \frac{q_0^*}{|q_0|^2}$, making $\QN$ into a normed division algebra.

An arbitrary two-qubit state can be written in the form
\begin{equation}
    \ket{\psi} = c_0\ket{00} + c_1\ket{01} + c_2\ket{10} + c_3\ket{11},
\end{equation}
with $|c_0|^2 + |c_1|^2 + |c_2|^2 + |c_3|^2 = 1$. This state is separable if $c_0c_3 = c_1c_2$.

Due to the normalization, each two-qubit state can be represented by a point on the 7-sphere $S^7$. We introduce a pair of quaternions
\begin{align*}
    q_0 &= c_0 +  c_1\qj\\
    q_1 &= c_2 +  c_3\qj,
\end{align*}
and define the quaternionic Pauli matrices
\begin{equation}
\begin{matrix*}[l]
    \sigma_1^q=\begin{pmatrix} 0 & 1 \\ 1 & 0 \end{pmatrix},
    &\sigma_2^q=\begin{pmatrix} 0 & -\qi \\ \qi & 0 \end{pmatrix},\\
    \sigma_3^q=\begin{pmatrix} 0 & -\qj \\ \qj & 0 \end{pmatrix},
    &\sigma_4^q=\begin{pmatrix} 0 & -\qk \\ \qk & 0 \end{pmatrix},\\
    \sigma_5^q=\begin{pmatrix} 1 & 0 \\ 0 & -1 \end{pmatrix}.
\end{matrix*}
\end{equation}

With these, we can define a quaternionic `Bloch vector' with coordinates 
\begin{equation*}
    x_i=\left\langle \sigma_i^q \right\rangle_\psi = \left(q_0^*,q_1^*\right)\sigma_i^q\begin{pmatrix} q_0 \\ q_1 \end{pmatrix}.
\end{equation*}

\subsection{$SL(2,\QN )$ and $5+1$-Dimensional Spacetime}\label{sub:quaspace}

Analogously to Sec.~\ref{sub:complspace}, we can write a point in $5+1$-dimensional Minkowski-space by means of the quaternion Hermitian matrix
\begin{equation}\label{eq:5+1dim}
    X = \frac{1}{2}\begin{pmatrix} x_0 + x_5 & x_1- x_2\qi- x_3\qj- x_4\qk \\ x_1+ x_2\qi+ x_3\qj+ x_4\qk & x_0-x_5 \end{pmatrix},
\end{equation}
whose determinant again gives the $5+1$-dimensional Minkowski interval, yielding $\mathrm{det}(X) = 0 \rightarrow x_0^2 - x_1^2 - x_2^2- x_3^2- x_4^2- x_5^2 = 0$. On these, the Lorentz transformations again act via conjugation by matrices from $SL(2,\QN)$. 

Due to the non-commutativity of the quaternions, some care has to be taken in defining this group. One possibility here is to take $SL(2,\QN)$ to be the group of $2\times2$ matrices $M_q$ with quaternionic entries,
\begin{equation}
  M_q = \begin{pmatrix} a & b \\ c & d \end{pmatrix}, a,b,c,d \in \QN,
\end{equation}
such that the Dieudonn\'e determinant $\Delta(M_q) = \left|ad - aca^{-1}b\right|$ is equal to $1$ \cite{venancio2020two}. Alternatively, one can construct the group via its Lie algebra $\mathfrak{sl}(2,\QN)$ \cite{baez2002octonions}.

We can now again, appealing to the Bloch representation, suppose each quaternionic qubit to parametrize a lightlike direction on the heavenly sphere, and obtain the restricted Lorentz group $SO^+(5,1)$. Hence, entangled two-qubit states give rise to the symmetries of $5+1$-dimensional Minkowski space \cite{brody2011six}.

Now suppose we single out a preferred complex structure on $\QN$, yielding a split $\QN \simeq \CN\oplus\CN$. Suppose we single out $\qi$ and thus, require that the Lorentz transformations commute with it. This will restrict $SL(2,\QN)$ to $SL(2,\CN)$, thus to the symmetries of $3+1$-dimensional Minkowski space. The extra spatial dimensions in this picture, $x_3$ and $x_4$, are due to the additional quaternionic imaginary directions $\qj$ and $\qk$, and hence, are ``ignored" by singling out a complex structure.

This is the first important point to note: if we want to represent quaternionic quantum mechanics within $3+1$-dimensional spacetime, we have to single out a preferred complex structure. Kaluza-Klein like dimensional reduction can thus be achieved by means of identifying a preferred copy of the complex plane within a higher division algebra. The motivation for such a singling out might then be given by the need to represent data obtained from such entangled states in the spacetime emerging from the area-law portion of the entanglement, with respect to which we might think of additional entanglement as introducing an `internal' geometry.

\subsection{Two Qubits and the Quaternionic Hopf Fibration}\label{sub:quahopf}

As in the complex case, we have above already implicitly introduced the second, quaternionic, Hopf fibration. To make it more explicit, we again start with a map from the two-qubit state space $S^7$ to the extended quaternionic plane,
\begin{equation}
    h_1^q: \begin{matrix*}[c] S^7 & \longrightarrow & \QN \cup\{\infty\} \\ (q_0,q_1) & \longrightarrow & q_0q_1^{-1} \end{matrix*},
\end{equation}
which is then composed with the inverse stereographic projection
\begin{equation}
    h_2^q: \begin{matrix*}[c] \QN \cup\{\infty\} & \longrightarrow & S^4 \\ q_0q_1^{-1} & \longrightarrow & (x_1,x_2,x_3,x_4,x_5) \end{matrix*},
\end{equation}
where again $\sum_{i=1}^5x_i^2=1$.

Thus, this shows how $S^7$ is realized as a nontrivial $S^3$-fibration over $S^4$,
\begin{equation}
    S^7 \stackrel{S^3}{\longrightarrow} S^4,
\end{equation}
exemplified in the fact that two points $(q_0,q_1)$ and $(q_0q,q_1q)$ where $|q|^2=1$ are mapped onto the same point of the base space $S^4$.

The most interesting aspect of this map is that it is entanglement sensitive \cite{mosseri2001geometry}: for each state with $c_0c_3 = c_1c_2$, i. e. for each separable state, we have
\begin{equation}
    x_3 = x_4 = 0 \Longrightarrow h_1(q_0,q_1)\in\CN\subset\QN.
\end{equation}
Thus, for non-entangled qubits, the fibration simplifies to $S^2\otimes S^2$, the product of the individual qubit spaces, and the extra dimensions associated with the additional quaternionic imaginary units vanish. Consequently, the matrix in Eq.~\eqref{eq:5+1dim}, parametrizing a point in $5+1$-dimensional Minkowski space, reduces to three spatial dimensions. This illustrates the role played by entanglement in the present construction: the entangled state of two qubits cannot be embedded within the $3+1$-dimensional spacetime derived from the symmetries of individual, separable qubits---there are additional degrees of freedom accounting for the fact that the states of the individual qubits alone do not fully fix the combined, entangled state. In the end, this merely signals the familiar fact that the reduced density matrices of the individual qubits do not fully characterize the state of an entangled system. It is the role and interpretation of these additional degrees of freedom that is the main focus of the present work.

Hence, only entangled qubit spaces lead to the above description proper. The fiber in the construction then contains the degrees of freedom of one qubit, while the base space contains the other plus the entanglement degrees of freedom. 

This also means the construction can be iterated: the $S^3$ fiber, as the state space of the primary qubit, can itself again be written as an $S^1$ fibration over an $S^2$ base.

\subsection{$Spin(5)$ and $SU(2)\times U(1)/\mathbb{Z}_2$}\label{sub:quaspin}

As in the discussion of the complex case (Sec.~\ref{sub:compspin}), following Ref.~\cite{krasnov2020so9} we consider matrices of the form
\begin{equation*}
    X(r,q)=\begin{pmatrix} r & L_q \\ L_{q^*} & -r \end{pmatrix},r\in\RN,q\in\QN,
\end{equation*}
where here, due to the non-commutative nature of the quaternions, the specification of left-multiplication is essential. These matrices now generate $Spin(5)$ in the case of $r^2 + |q|^2 = 1$.

As in the complex single-qubit case, $Spin(5)$ can be understood as the equivariance group of the quaternionic Hopf fibration \cite{gluck1986geometry}, where again the $Spin(5)$-action on the 4-sphere $S^4$ is induced by its action on $\RN^5$, and its action on the 7-sphere $S^7$ follows from the isomorphism $Spin(5)\simeq Sp(2) = U(2,\QN)$ by the left action of $U(2,\QN)$ on $\QN^2$.

As before, we can write the Hopf fibration in terms of coset spaces to make this more explicit, yielding \cite{hatsuda2009coset}
\begin{equation}
 S^7\stackrel{S^3}{\longrightarrow}S^4 \equiv \frac{Spin(5)}{Spin(3)}\stackrel{\frac{Spin(4)}{Spin(3)}}{\longrightarrow}\frac{Spin(5)}{Spin(4)}.
\end{equation}

It can now be shown that the subgroup of $Spin(5)$ preserving the split $\QN\simeq\CN\oplus\CN$ is equal to $SU(2)\times U(1)/\mathbb{Z}_2$. To extend to the octonionic case, the argument proceeds on the Lie algebra level. We will not review this argument in its entirety here, and instead refer to the presentation in Ref.~\cite{krasnov2020so9}. 

However, thanks to the associativity of the quaternions, there exists a way to cast the argument in matrix terms, which we reproduce here in summary to build some intuition for the general claim.

First, note that we can represent quaternions as $2\times 2$ unitary matrices, 
\begin{equation}
    Q = \begin{pmatrix} z & u \\ -u^* & z^* \end{pmatrix},
\end{equation}
with $z,u\in\CN$. Quaternion conjugation then reduces to Hermitean conjugation, $Q^* = Q^\dagger$, and the quaternionic norm is given by $|Q|^2=\frac{1}{2}\mathrm{Tr}\left(QQ^\dagger\right)=\mathrm{det}(Q)$.

$Spin(5)$ can then, via the exceptional isomorphism with $U(2,\QN)$, be represented by $2\times2$ matrices with quaternionic entries,
\begin{equation}
    g = \begin{pmatrix} A & B \\ C & D \end{pmatrix},
\end{equation}
with $A$, $B$, $C$ and $D$ being quaternions, viewed as $2\times2$ matrices. These act on quaternionic column vectors 
\begin{equation}
    S = \begin{pmatrix} Q \\ P \end{pmatrix} = \begin{pmatrix} z & u \\ -u^* & z^* \\ v & w \\ -v^* & w^* \end{pmatrix}\in\QN^2
\end{equation}

Now, as before, fix a quaternion imaginary direction $\qi$. Then, the quaternions commuting with this choice are those whose $\qj$, $\qk$-components vanish. In terms of the representation as $2\times2$-matrices, these are the diagonal matrices. Consequently, we are left with the subgroup of $U(2,\QN)$ where $A$, $B$, $C$ and $D$ are diagonal matrices satisfying constraints imposed by the requirement of preserving the square norm \cite{krasnov2020so9}, which can be parametrized as follows:
\begin{align*}
    A=\begin{pmatrix}e^{i\phi}a & 0 \\ 0 & e^{-i\phi}a^* \end{pmatrix}, & B =\begin{pmatrix}e^{i\phi}b & 0 \\ 0 & e^{-i\phi}b^* \end{pmatrix} \\
    C=-\begin{pmatrix}e^{i\phi}b^* & 0 \\ 0 & e^{-i\phi}b \end{pmatrix}, & D =\begin{pmatrix}e^{i\phi}a^* & 0 \\ 0 & e^{-i\phi}a \end{pmatrix}
\end{align*}
The action on $S$ then induces the action on the complex subspaces given by
\begin{equation}\label{eq:SU2xU1}
    \begin{matrix*}[l]
    \begin{pmatrix} z \\ w \end{pmatrix}&\longrightarrow e^{i\phi}\begin{pmatrix} a & b \\ -b^* & a^* \end{pmatrix}\begin{pmatrix} z \\ w \end{pmatrix}\\
    \begin{pmatrix} u \\ v \end{pmatrix}&\longrightarrow e^{i\phi}\begin{pmatrix} a & b \\ -b^* & a^* \end{pmatrix}\begin{pmatrix} u \\ v \end{pmatrix},
    \end{matrix*}
\end{equation}
thus showing that the subgroup preserving the split is just $SU(2)\times U(1)/\mathbb{Z}_2$. This has, in fact, a nicely intuitive interpretation: the $SU(2)\simeq SO(3)$ transformation acts on the $\{x_1,x_2,x_5\}$-coordinates of the base space, while the $U(1)\simeq SO(2)$ permutes the $\{x_3,x_4\}$-directions. Hence, the splitting of $\QN$ into two copies of $\CN$ has the effect that the transformations `mixing' the two sets of coordinates are removed, leaving only those associated with each complex subspace individually.

It should be noted that we are engaging in some double-counting of degrees of freedom, here: the $\{x_1,x_2,x_5\}$-coordinates that are taken to parametrize the celestial sphere in Sec.~\ref{sub:quaspace} is here taken to be subject to the $SU(2)$-symmetry. This will be addressed in Sec.~\ref{sec:mass}.

\section{The Octonionic Case: Three Qubits}\label{sec:oct}

The third step in this construction takes us to the octonionic case. Remarkably, this is also the final step: the octonions are the largest division algebra, and the octonionic one is the final Hopf fibration. The next algebra that can be constructed via the Cayley-Dickson process, the sedenions, is no longer a division algebra, as it possesses nontrivial zero divisors, and hence, the Hopf map cannot always be defined.

There is thus a certain finality to this case. Having this construction naturally stop here then may go some way towards explaining why it is just this case---and not any one further `up the ladder'---that might play a role in nature. More complex states then should not be expected to bring anything fundamentally new to the table---or at least, nothing that yields novel gauge symmetries in the way described here.

We will not give an in-depth introduction to the octonion algebra here, instead referring the reader to Baez' magnificent exposition of their manifold subtle, and sometimes surprising, connections with the broader mathematical landscape \cite{baez2002octonions}. Suffice it to say, here, that octonions can be generated by taking a pair of quaternions and introducing an additional imaginary unit, $e_4$ (where we identify $(\qi,\qj,\qk) \equiv (e_1,e_2,e_3)$), yielding (following the conventions of \cite{bernevig2003geometry})
\begin{align*}
    o_0 &= q_0 + q_1e_4 \\
        &= \underbrace{r_0 + r_1e_1 + (r_2 + r_3e_1)e_2}_{q_0} + \underbrace{\left[r_4 + r_7e_1 + (r_6 - r_5e_1)e_2\right]}_{q_1}e_4.
\end{align*}
This gives the quaternionic triples
\begin{equation*}
    (123),(246),(435),(367),(651),(572),(714). 
\end{equation*}
A salient property of the octonions is their non-associative nature: $(e_ie_j)e_k = -e_i(e_je_k)$ for $i\ne j \ne k$ and $e_ie_j\ne\pm e_k$. Octonions are, however, still \emph{alternative}: for any two elements $o_1,o_2\in\ON$, it holds that
\begin{equation*}
    (o_1o_1)o_2 = o_1(o_1o_2),\: (o_1o_2)o_1 = o_1(o_2o_1),\: (o_2o_1)o_1 = o_2(o_1o_1).
\end{equation*}

Octonion conjugation takes $o_0$ to $o_0^*$ with
\begin{equation*}
    o_0^* = r_0 - \sum_{i=1}^7 r_ie_i,
\end{equation*}
which again lets us define a norm $|o_0|^2 = o_0o_0^* = \sum_{i=0}^8r_i^2$ and a multiplicative inverse $o_0^{-1}=\frac{o_0^*}{|o_0|^2}$.

An arbitrary three-qubit state can be written as
\begin{equation}
    \begin{matrix*}[c]
        \ket{\psi} = c_0\ket{000} + c_1\ket{001} + c_2\ket{010} + c_3\ket{011} \\ + c_4\ket{100} + c_5\ket{101} + c_6\ket{110} + c_7\ket{111},
    \end{matrix*}
\end{equation}
with $\sum_{i=0}^8|c_i|^2 = 1$. The normalization condition entails that a three-qubit state can be represented by a point on the 15-sphere $S^{15}$. We now define the following four quaternions:
\begin{equation}
    \begin{matrix*}[l]
        q_0 = c_0 + c_1e_2, & q_1 = c_2 + c_3^*e_2 \\
        q_2 = c_4 + c_5e_2, & q_3 = c_6 + c_7^*e_2,
    \end{matrix*}
\end{equation}
which we combine into the octonions
\begin{equation}
    \begin{matrix*}[l]
        o_0 = q_1 + q_2e_4\\
        o_1 = q_3 + q_4e_4.
    \end{matrix*}
\end{equation}
This particular choice avoids introducing an anisotropy on $S^{15}$ \cite{bernevig2003geometry}.

We can now introduce an octonionic analogue of the Pauli matrices
\begin{align*}
    \sigma_1^o & = \begin{pmatrix} 0 & 1 \\ 1 & 0 \end{pmatrix}\\
    \sigma_i^o & = \begin{pmatrix} 0 & -e_{i-1} \\ e_{i-1} & 0 \end{pmatrix}\\
    \sigma_9^o & = \begin{pmatrix} 1 & 0 \\ 0 & -1 \end{pmatrix},
\end{align*}
where $i$ runs from $2$ to $8$. This allows us to define an octonionic `Bloch vector', whose coordinates are
\begin{equation*}
    x_i=\left\langle \sigma_i^o \right\rangle_\psi = \left(o_0^*,o_1^*\right)\sigma_i^o\begin{pmatrix} o_0 \\ o_1 \end{pmatrix},
\end{equation*}
with $i$ here running from $1$ to $9$. 

\subsection{`$SL(2,\ON )$' and $9+1$-Dimensional Spacetime}\label{sub:octspace}

Despite the nonassociativity of the octonions, the general picture described above is still valid. As before, we can take matrices of the form
\begin{equation}
    X = \frac{1}{2}\begin{pmatrix} x_0 + x_9 & x_1- \sum_{i=1}^7x_{i+1}e_i \\ x_1+ \sum_{i=1}^7x_{i+1}e_i & x_0-x_9 \end{pmatrix}
\end{equation}
to parametrize the light-cone in $9+1$-dimensional Minkowski space (recall that such matrices always have a well-defined determinant, Eq.~\eqref{eq:divdet}), on which the Lorentz transformations are realized via conjugation by matrices from `$SL(2,\ON)$'. 
However, the octonions being non-associative in addition to non-commutative, it is not immediately clear what, exactly, could be meant by this group. Nevertheless, it can be shown that a construction exists that generalizes the complex and quaternionic case, and again allows us to think of (pure) octonionic `qubits' as parametrizing the $9+1$-dimensional celestial sphere \cite{manogue1993finite} (s. a. \cite{dray2009octonionic,baez2002octonions}).

\subsection{Three Qubits and the Octonionic Hopf Fibration}\label{sub:octhopf}

Due to the normalization $|o_0|^2 + |o_1|^2 = 1$, an octonionic qubit defines a point on the sphere $S^{15}$. There again exists a map taking points of the sphere to the extended octonionic plane,
\begin{equation}
    h_1^o: \begin{matrix*}[c] S^{15} & \longrightarrow & \ON \cup\{\infty\} \\ (o_0,o_1) & \longrightarrow & o_0o_1^{-1} \end{matrix*},
\end{equation}
which can then be projected to $S^8$ by means of
\begin{equation}
    h_2^o: \begin{matrix*}[c] \ON \cup\{\infty\} & \longrightarrow & S^8 \\ o_0o_1^{-1} & \longrightarrow & \left(x_i\right)_{i=1\ldots9} \end{matrix*},
\end{equation}
once more with $\sum_{i=1}^9x_i^2=1$. As before, this is just the map taking $\left(o_0,o_1\right)$ to the `Bloch-vector' coordinates,
\begin{equation}
    h_2\circ h_1\left(o_0,o_1\right) = \left(x_i\right)_{i=1\ldots9}.
\end{equation}

This realizes $S^{15}$ as an $S^7$-fibration over an $S^8$ base, 
\begin{equation}
    S^{15} \stackrel{S^7}{\longrightarrow} S^8,
\end{equation}
and as in the previous case, the map reduces to a projection into the complex subspace of $\ON$ in case the state is separable into a single-qubit and two-qubit state. Hence, the base space $S^8$ contains the (projective) state of one qubit, along with entanglement degrees of freedom, while the fiber contains the state of the leftover two qubits.

\subsection{$Spin(9)$ and $SU(3)\times SU(2)\times U(1)/\mathbb{Z}_6$}\label{sub:octspin}

As with the previous two sections, there exists a generalization of the connection between the spin-groups and Hopf fibrations to the octonionic case. One can arrive at an analogous description of the group $Spin(9)$ in terms of matrices of the form given in Eq.~\eqref{eq:spinmat}, with $L_c$ replaced by $L_o$, denoting left-multiplication by octonions. However, we can also again simply note that the octonionic Hopf fibration can be equivalently given as \cite{ornea2013spin}
\begin{equation}
 S^{15}\stackrel{S^7}{\longrightarrow}S^8 \equiv \frac{Spin(9)}{Spin(7)}\stackrel{\frac{Spin(8)}{Spin(7)}}{\longrightarrow}\frac{Spin(9)}{Spin(8)}.
\end{equation}

Then, one must find the subgroup of $Spin(9)$ respecting a split $\ON\simeq\CN\oplus\CN^3$, that is, singling out an octonion imaginary direction. It was shown by Dubois-Violette and Todorov \cite{dubois2019exceptional} (see also \cite{todorov2019exceptional,dubois2016exceptional,todorov2018deducing}), and put into a more explicit form by Krasnov \cite{krasnov2020so9}, that this is just the gauge group of the Standard Model, $SU(3)\times SU(2)\times U(1)/\mathbb{Z}_6$.

We will not give the full argument here, and refer to the cited literature instead. However, as in the quaternionic case, there is an intuitive way of viewing the group's action on the base space of the Hopf fibration. Under the split $\ON\simeq\CN\oplus\CN^3$, the coordinates $x_i$ split into two sets, $\left\{x_1,x_2,x_9\right\}$ from the singled-out complex subspace, and $\left\{x_3,x_4,x_5,x_6,x_7,x_8\right\}$ from the additional octonionic complex directions (the entanglement degrees of freedom). Here, the first set characterizes one qubit, while the second gives the entanglement with the other two. Thus, we must look at the transformations that do not mix these sets.

As before, the triplet $\left\{x_1,x_2,x_9\right\}$ is acted on by rotations from $SO(3)\simeq SU(2)$. We now use the embedding of $U(1)\times SU(3)$ into $SU(4)$, given by
\begin{equation}\label{eq:embed}
    \left(\alpha,g\right) \longrightarrow \begin{pmatrix}
    \alpha^{-3} & 0 \\ 0 & \alpha g,
    \end{pmatrix}
\end{equation}
where $\alpha\in U(1)$ and $g\in SU(3)$ \cite{krasnov2020so9}. With $SU(4)\simeq Spin(6)$, this then acts on the six remaining directions, as required.

Finally, from the above embedding, one sees that the $\ON^2$-spinor splits into two representations, with the single factor of $\CN$ being associated to an $SU(3)$-singlet of $U(1)$-charge $-1$, and the $\CN^3$ associated to a $SU(3)$ triplet of $U(1)$-charge $\frac{1}{3}$, both transforming as a doublet under $SU(2)$ \cite{krasnov2020so9}. Consequently, the octonionic qubit can be seen to contain precisely one generation of left-handed particles with the quantum numbers of the Standard Model, after being split into a leptonic part $L$ from the factor $\CN$ and a quark part $Q$ from $\CN^3$:
\begin{equation}
    \left.
    \begin{matrix}
        L &\longrightarrow &\alpha^{-3}hL \\
        Q &\longrightarrow &\alpha g h Q\\
    \end{matrix}
    \right\} (g,h,\alpha)\in SU(3)\times SU(2) \times U(1)
\end{equation}

\subsection{Summary}

We have outlined a correspondence between the symmetries of few-qubit states and the gauge symmetries of the standard model. To help intuition, it is useful to recapitulate the construction in a pictorial way, highlighting the action of the symmetry groups.

\begin{figure}[h] 
 \centering
 \begin{overpic}[width=0.7\textwidth]{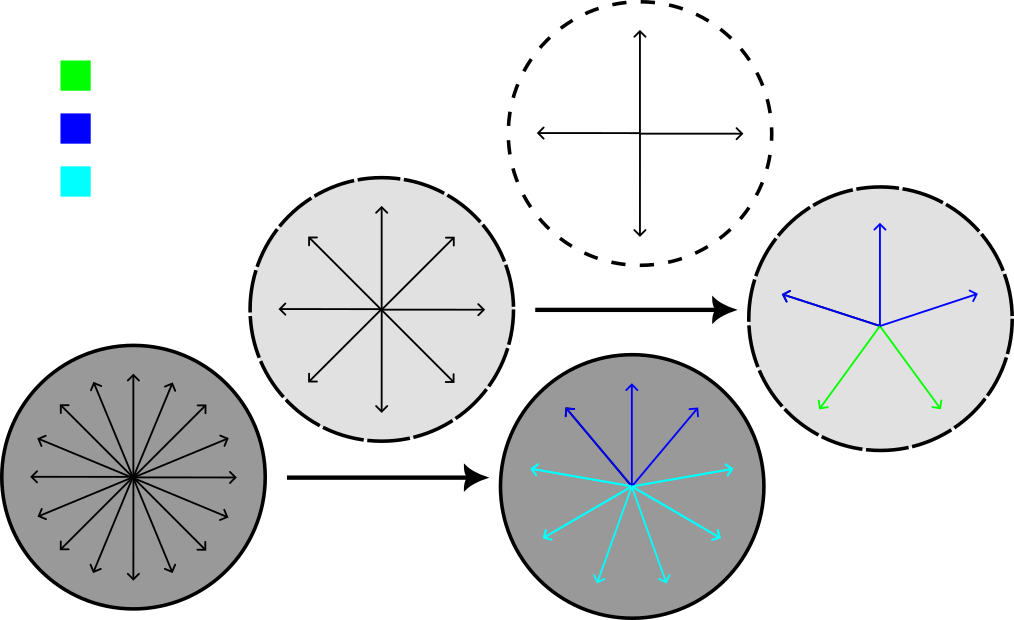}
  \put(10,52.3){$U(1)$}
  \put(10,47){$SU(2)$}
  \put(10,42){$SU(3)\times U(1)$}
  \put(22,2){$S^{15}$}
  \put(72,2){$S^{8}$}
  \put(45.5,18){$S^{7}$}
  \put(95.5,18){$S^{4}$}
  \put(72,36){$S^{3}$}
 \end{overpic}
\caption{The iterated Hopf fibration, starting out with $S^{15}$ containing the 3-qubit state as an $S^{7}$ fibration over $S^{8}$, with the $S^7$ fiber itself being an $S^3$-fiber over $S^4$, and similarly for the $S^3$-fiber again (not shown explicitly). The colors highlight the coordinates permuted by the various symmetry groups.}
\label{pic:FibIter}
\end{figure}
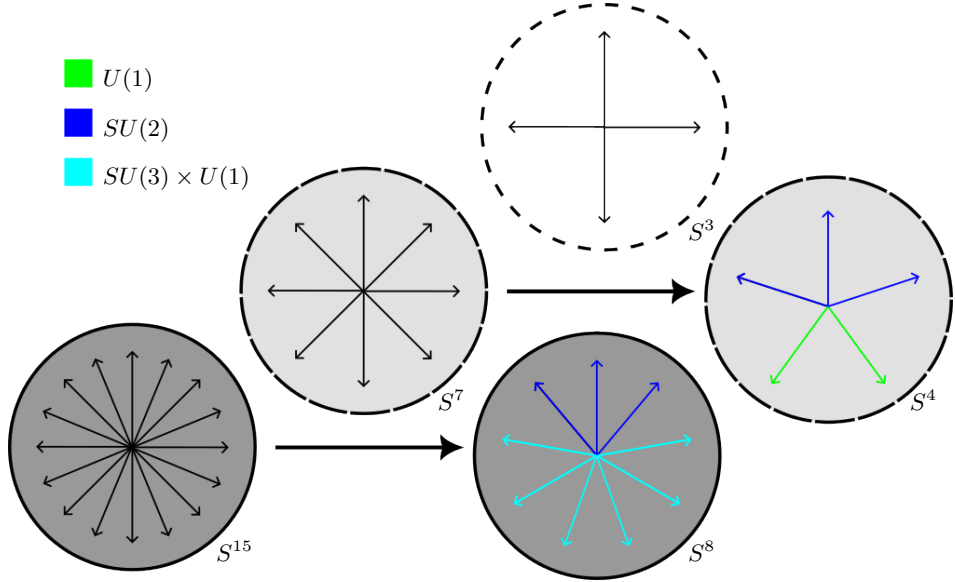

In Fig.~\ref{pic:FibIter}, the iteration of the Hopf fibration, starting with the $S^{15}$ containing the three qubit state, is depicted. Arrows indicate the coordinate axes of each space. At each stage, the colors in the fibration's base indicate the actions of the symmetry group: within the $S^8$ base, the $SU(3)\times U(1)$ symmetry acts on the directions $\{x_3,\ldots,x_8\}$, while the $SU(2)$ permutes the remaining three coordinates. On the next level, the directions $\{x_1,x_2,x_5\}$ are acted on by elements of $SU(2)$, with $U(1)$ yielding rotations in the subplane spanned by the other two directions.

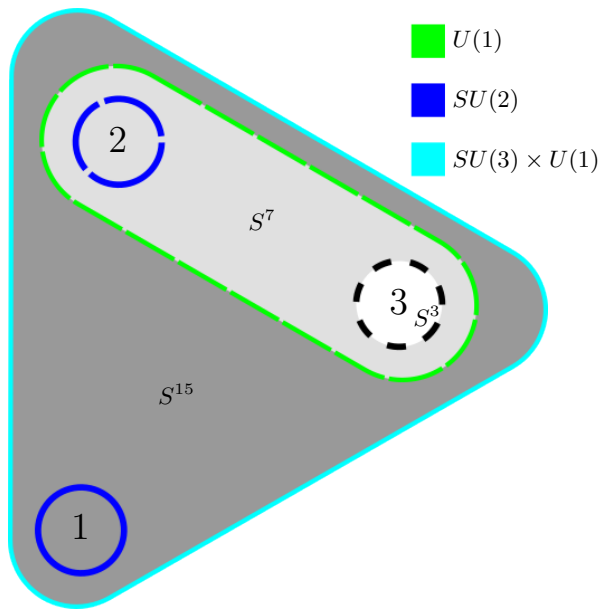
\begin{figure}[h] 
 \centering
 \begin{overpic}[width=0.4\textwidth]{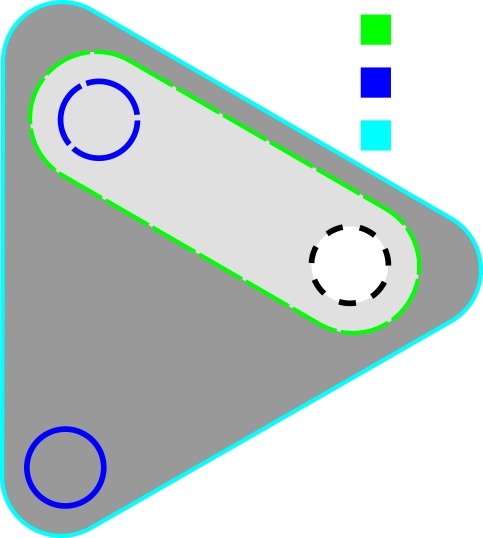}
  \put(10.5,11.5){\Large $1$}
  \put(16.7,76){\Large $2$}
  \put(63.5,49){\Large $3$}
  \put(74,93.5){$U(1)$}
  \put(74,83.5){$SU(2)$}
  \put(74,73.5){$SU(3)\times U(1)$}
  \put(25,34){$S^{15}$}
  \put(40,63){$S^{7}$}
  \put(67.5,47){$S^{3}$}
 \end{overpic}
\caption{The 3-qubit state with the association between qubit- and entanglement degrees of freedom and the respective action of the symmetry groups highlighted.}
\label{pic:3Qubit}
\end{figure}

Fig.~\ref{pic:3Qubit} shows the three qubit state, together with the qubit- and entanglement degrees of freedom, highlighting again the action of each symmetry. The dark grey patch represents $S^{15}$, containing one qubit and the degrees of freedom corresponding to its entanglement with the other two qubits in the base, and the other two qubits in the fiber. The $SU(2)$-symmetry acts on the coordinates containing the qubit, with the $SU(3)\times U(1)$ acting on the coordinates parametrizing the entanglement. Likewise, the $S^7$ fiber contains in its base the second qubit, on which again $SU(2)$ transformations act, and the entanglement of that qubit to the third one, that is itself subject to a $U(1)$-action. 

\subsection{Further Development}

As noted, the construction above terminates in a natural way with the octonionic case. However, of course, the universe is not limited to three qubits and their entanglement. Therefore, investigation into the properties of state spaces beyond three qubits seems a likely avenue for further developing the model. 

Moving to four qubits, it is then tempting to appeal to the next algebra in the Cayley-Dickson sequence, the sedenions $\SN$. Of course, because of the existence of zero divisors, the Hopf map cannot be defined in a straightforward way. Hence, a fully analogous continuation of the sequence above does not seem feasible---which indeed may be an indication of the apparent relevance of this case in nature.

Nevertheless, it has been proposed that the sedenions could be utilized to quantify four-qubit entanglement in a similar way, with a single four-qubit state being a point on $S^{31}$ \cite{pinilla2009hopf}. Furthermore, the discrete $S_3$ automorphism of $\SN$ has been argued to yield a natural realization of the triple-generational structure of the Standard Model \cite{gresnigt2023three, gourlay2024algebraic, tang2023unified}, providing essentially three copies of the octonion algebra. 

Hence, while, as expected, no essentially new phenomenology (e.g. fifth forces) becomes apparent through enlarging the model, it seems nevertheless possible to capture further relevant structure of the Standard Model, such as most notably the generational structure.

\section{Entanglement and Massive Particles}\label{sec:mass}

The above has given us a way to interpret the degrees of freedom contained within a single octonionic qubit as the \emph{left-handed} part of one generation of the fermion content of the Standard Model, transforming under the gauge symmetry $SU(3)\times SU(2)\times U(1)/\mathbb{Z}_6$ in the appropriate way, after singling out a complex direction and splitting $\ON$ into $\CN\oplus\CN^3$. 

However, some ambiguity still exists about the coordinates $\left\{x_1,x_2,x_9\right\}$: if we view the octonionic spinor as spanning the $9$-dimensional heavenly sphere, these coordinates should be interpreted as the spatial coordinates upon dimensional reduction by means of singling out a special complex direction; yet, they are acted upon by the $SU(2)$-part of the Standard Model symmetry. 

What comes to our aid is the observation that only half the particle content---the left-handed part---of the Standard Model is subject to the $SU(2)$-symmetry: the Standard Model is a chiral theory. Hence, if we instead view $\left\{x_1,x_2,x_9\right\}$ as spanning the $3$-dimensional celestial sphere, only the $U(1)\times SU(3)$-symmetry remains, and the $\ON^2$-spinor models a right-handed generation instead.

It then seems suggestive to move on to $\ON^4\simeq\CN^{16}$, which is indeed large enough to accommodate a full generation of Standard Model fermions. 

But let us first step back and observe that, so far, we have only considered pure states. These, as discussed above, give rise to light-like vectors (in $3+1$, $5+1$, and $9+1$ dimensions respectively). Hence, these can, at best, model massless particles propagating at the speed of light. However, we can appeal here to an observation due to Kiosses \cite{kiosses2014quantum} and Morikoshi \cite{morikoshi2015entanglement}: while a pure density matrix will have a Bloch vector of norm $1$, a mixed-state density matrix defines a point at the interior of the Bloch sphere---and hence, in the correspondence above, a time-like direction within the light-cone, in other words, the propagation of a massive particle.

Let us for the moment consider a single, unnormalized spinor
\begin{equation}
    \ket{\psi_A} = \begin{pmatrix} a \\ b \end{pmatrix},
\end{equation}
such that $a,b\in\CN$, $|a|^2 + |b|^2 = k_0$. Its density matrix is
\begin{equation}
    \rho_\psi = \frac{1}{2}\begin{pmatrix}
     k_0^A + k_3^A & k_1^A-ik_2^A \\ k_1^A+ik_2^A & k_0^A - k_3^A
     \end{pmatrix},
\end{equation}
which yields as before
\begin{equation}\label{eq:detrho}
    \mathrm{det}(\rho) = (k_0^A)^2 - (k_1^A)^2 - (k_2^A)^2 - (k_3^A)^2 = 0.
\end{equation}

We can find a linearization of this equation of the form
\begin{equation}\label{eq:Weyl}
    \left(k_0^A - k_i^A\sigma^i\right)\ket{\psi_A} = 0,
\end{equation}
where the repeated index is summed over. 

Acting on this from the left with $\bra{\psi}$ recovers Eq.~\eqref{eq:detrho}. This has the form of a Weyl-equation in the momentum representation; hence, we have renamed our coordinates $x_i\rightarrow k_i$ to illustrate this interpretation.

Explicitly, the $k_i$ are:
\begin{equation}\label{eq:kcomp}
    \begin{matrix*}[l]
    k_0^A = |a|^2 + |b|^2 \\
    k_1^A = a^*b + b^*a \\
    k_2^A = -i(a^*b - b^*a) \\
    k_3^A = |a|^2 - |b|^2
    \end{matrix*}
\end{equation}

Consider now a general two-qubit state of the form
\begin{equation}
    \ket{\Uppsi} = a\ket{00} + b\ket{01} + c\ket{10} + d\ket{11},
\end{equation}
with $\langle\Uppsi|\Uppsi\rangle = K_0$. 

Then let us define the matrices $\Sigma_i$:
\begin{align}
    \nonumber
    &\Sigma_1 = \begin{pmatrix}
     0 & -1 & 0 & 0 \\
    -1 & 0 & 0 & 0 \\
     0 & 0 & 0 & 1 \\
     0 & 0 & 1 & 0 \\
    \end{pmatrix}, 
    &\Sigma_2 = \begin{pmatrix}
     0 & i & 0 & 0 \\
     -i & 0 & 0 & 0 \\
     0 & 0 & 0 & -i \\
     0 & 0 & i & 0 \\
    \end{pmatrix}, \\
    &\Sigma_3 = \begin{pmatrix}
     0 & 0 & 1 & 0 \\
     0 & 0 & 0 & 1 \\
     1 & 0 & 0 & 0 \\
     0 & 1 & 0 & 0 \\
    \end{pmatrix}, 
    &\Sigma_4 = \begin{pmatrix}
     0 & 0 & -i & 0 \\
     0 & 0 & 0 & -i \\
     i & 0 & 0 & 0 \\
     0 & i & 0 & 0 \\
    \end{pmatrix}, \\
    \nonumber
    &\Sigma_5 = \begin{pmatrix}
     -1 & 0 & 0 & 0 \\
     0 & 1 & 0 & 0 \\
     0 & 0 & 1 & 0 \\
     0 & 0 & 0 & -1 \\
    \end{pmatrix}.
\end{align}

With these, we have again
\begin{equation}
    \left(K_0-K_i\Sigma^i\right)\ket{\Uppsi} = 0
\end{equation}
for
\begin{equation}
    \begin{matrix*}[l]
        K_0 = \ex{\Uppsi|\Uppsi} = |a|^2 + |b|^2 + |c|^2 + |d|^2 = k_0^A + k_0^B\\
        K_1 = \ex{\Uppsi|\Sigma_1|\Uppsi} = -a^*b-b^*a+c^*d+d*c = -k_1^A + k_0^B\\
        K_2 = \ex{\Uppsi|\Sigma_2|\Uppsi} = ia^*b-ib^*a - ic^*d + id^*c = -k_2^A + k_2^B\\
        K_3 = \ex{\Uppsi|\Sigma_3|\Uppsi} = a^*c +c^*a +b^*d +d^*b = \ex{\psi_A|\psi_B} + \ex{\psi_B|\psi_A}\\
        K_4 = \ex{\Uppsi|\Sigma_4|\Uppsi} = i c^*a + i d^*b -ia^*c -ib^*d = -i\ex{\psi_A|\psi_B} + i\ex{\psi_B|\psi_A}\\
        K_5 = \ex{\Uppsi|\Sigma_5|\Uppsi} = -|a|^2 + |b|^2 + |c|^2 - |d|^2 = -k_3^A + k_3^B
    \end{matrix*}
\end{equation}
Here, the $k_i^B$ are the Bloch vector coordinates of a further (unnormalized) spinor $\ket{\psi_B}=(c,d)^T$. With this, the four-spinor
\begin{equation}
    \ket{\Uppsi}=\ket{\psi_A}\oplus\ket{\psi_B}=\begin{pmatrix}
        a\\b\\c\\d
    \end{pmatrix}
\end{equation}
defines a light ray in $5+1$ dimensions. 

We can then write in block-matrix form
\begin{equation}
    \begin{pmatrix}
        K_0\mathds{1}_2-K_1\sigma^1-K_2\sigma^2-K_3\sigma^3 & K_3+iK_4\mathds{1}_2 \\
        K_3-iK_4\mathds{1}_2 & K_0\mathds{1}_2+K_1\sigma^1+K_2\sigma^2+K_3\sigma^3
    \end{pmatrix}
    \begin{pmatrix}
        \ket{\psi_A}\\\ket{\psi_B}
    \end{pmatrix}=0
\end{equation}

If we now introduce 
\begin{equation}\label{eq:mass}
     me^{\pm i\phi} = K_3\mp iK_4,
\end{equation}
we obtain, 
\begin{equation}
     \begin{matrix*}[l]
        \left(K_0 - K_1\sigma_1 - K_2\sigma_2 - K_5\sigma_3 \right)\ket{\psi_A} = me^{-i\phi}\ket{\psi_B}, \\
        \left(K_0 + K_1\sigma_1 + K_2\sigma_2 + K_5\sigma_3 \right)\ket{\psi_B} = me^{i\phi}\ket{\psi_A},
    \end{matrix*}
\end{equation}
or, after multiplication of the first with $e^{i\phi}$ \cite{kiosses2014quantum},
\begin{equation}\label{eq:emgauge}
     \begin{matrix*}[l]
        \left(K_0 - K_1\sigma_1 - K_2\sigma_2 - K_5\sigma_3 \right)e^{i\phi}\ket{\psi_A} = m\ket{\psi_B}, \\
        \left(K_0 + K_1\sigma_1 + K_2\sigma_2 + K_5\sigma_3 \right)\ket{\psi_B} = me^{i\phi}\ket{\psi_A},
    \end{matrix*}
\end{equation}
where we see that $\ket{\psi_A}$ picks up a $U(1)$ phase factor from the rotation through the circle $m^2 = K_3^2 + K_4^2$ by an angle $\phi$. This, we recall, is the $U(1)$ of the $SU(2)\times U(1)$-subgroup of $Spin(5)$ respecting the split $\QN \simeq \CN \oplus \CN$. If we thus want to keep this as a symmetry of the theory, we should require that it remains invariant under such transformations. Consequently, we see that the symmetry preserving the split indeed acts as a gauge transformation $\ket{\psi_A}\rightarrow e^{i\phi}\ket{\psi_A}$, producing a chiral $U(1)$ gauge theory.

This identification of the $U(1)$ phase with electric charge is also made by Budinich \cite{budinich2002geometry}, where the above equation is interpreted as describing the proton and the neutron.

Absorbing the phase factor thus into $\ket{\psi_A}$, this can be brought into a more standard form by introducing
\begin{equation}
    \ket{\Uppsi^\prime}=\begin{pmatrix}\ket{\psi_B} \\ \ket{\psi_A} \end{pmatrix}
\end{equation}
together with the matrices $\gamma_i$ in the chiral representation
\begin{equation}
    \begin{matrix*}[l]
        \gamma_0 = \begin{pmatrix}
                0 & \mathds{1}_2 \\ \mathds{1}_2 & 0
            \end{pmatrix}, &
        \gamma_i = \begin{pmatrix}
                0 & \sigma_i \\ -\sigma_i & 0
            \end{pmatrix},
    \end{matrix*}
\end{equation}
which finally yields the Dirac equation
\begin{equation}\label{eq:Dirac}
    \left(p_i\gamma^i - m\right)\ket{\Uppsi^\prime} = 0,
\end{equation}
where we have for ease of notation relabeled $K_5 \rightarrow p_3$, and otherwise $K_i \rightarrow p_i$. Note that, to recover $K_0^2 - \sum_iK_i^2 = 0$, we must act on this from the left with $\bra{\widetilde{\Uppsi}^\prime_{AB}} = \ket{\Uppsi^\prime_{AB}}^\dagger\gamma_0$.

Now introducing
\begin{equation}
    \gamma_5 = i\gamma_0\gamma_1\gamma_2\gamma_3 = \begin{pmatrix}
        -\mathds{1}_2 & 0 \\ 0 & \mathds{1}_2
    \end{pmatrix}
\end{equation}
and the left- and right-chiral projectors
\begin{equation}
    \begin{matrix*}[l]
        P_L = \frac{1}{2}(\mathds{1}_4 - \gamma_5) = \begin{pmatrix}
            \mathds{1}_2 & 0 \\ 0 & 0
        \end{pmatrix}, \\
        P_R = \frac{1}{2}(\mathds{1}_4 + \gamma_5) = \begin{pmatrix}
            0 & 0 \\ 0 & \mathds{1}_2
        \end{pmatrix},
    \end{matrix*}
\end{equation}

we identify the left- and right-chiral components, $\ket{\psi_L} = P_L\ket{\Uppsi^\prime} = \ket{\psi_B}$ and $\ket{\psi_R} = P_R\ket{\Uppsi^\prime} = \ket{\psi_A}$. Hence, the entangled two-qubit state can be interpreted as a left- and a right-handed spinor in $3+1$ dimensions of mass $m$, with $m^2 = K_3^2 + K_4^2 = K_0^2 - K_1^2 - K_2^2 - K_5^2$. For vanishing entanglement, we again obtain two separate instances of $S^2$, parametrizing two points, two lightrays on the celestial sphere in three dimensional space.

Consequently, we have two different interpretations of the two-qubit state: in the interpretation given in Sec.~\ref{sec:quat}, we obtain the internal symmetries of a left-handed fermion generation (without color degrees of freedom, i. e. $SU(2)\times U(1)/\mathbb{Z}_2$), while the interpretation in this section yields the spatial degrees of freedom of a left- and a right-handed fermion, together with the internal $U(1)$ of the right-handed part. 

Putting both together, then, yields all the right ingredients for a unified description of a full generation (again, without the color symmetry). As $\QN^4\simeq \CN^8$, this means that we need a three-qubit state, however, now interpreted in terms of two copies of the construction referring to the quaternionic Hopf fibration. 

This then suggests an extension of these observations to the octonionic case. The move from $\ON^2$ to $\ON^4$ is equivalent to moving from a three-qubit to a four-qubit state. This gives rise to two copies of the octonionic Hopf fibration---which may be motivated by the fact that there is no fourth fibration, and thus, the construction as presented in this paper finds its natural endpoint there. 

As discussed in Sec.~\ref{sub:octspin}, the coordinates after singling out a complex direction split into two sets, $\{K_0,K_1,K_2,K_9\}$ and $\{K_3,K_4,K_5,K_6,K_7,K_8\}$ in the notation of this section, and with the norm $K_0$ made explicit. These obey
\begin{equation}
    K_0^2-K_1^2-K_2^2-K_9^2 = K_3^2+K_4^2+K_5^2+K_6^2+K_7^2+K_8^2,
\end{equation}
where the right side (giving the `mass term' as in Eq.~\eqref{eq:mass}) is acted on by the $SO(6)\simeq SU(4)$ into which the $SU(3)\times U(1)$-part of the Standard Model symmetry is embedded, as in Eq.~\eqref{eq:embed}. Thus, we see that the octonionic qubit can be interpreted, analogously to the quaternionic one, as the left-handed part of one family of fermions transforming under the full Standard Model symmetry $SU(3)\times SU(2) \times U(1)/\mathbb{Z}_6$, or as a right-handed part, together with the energy-momentum four-vectors of the left- and right-handed fermion, transforming under the residual $SU(3)\times U(1)$-symmetry. Putting both together, then, seems well poised to yield the description of a full generation of fermions. However, we leave the full formulation of the model for future work.

We note here that in this form, the model is also closely related to that proposed by Budinich \cite{budinich2002geometry}. 

\section{Quantum Simulation of Gauge Fields}\label{sec:Simul}

The correspondence between simple systems of few qubits and spacetime geometries has been found to provide fertile ground for implementing spacetime analog systems on present-day quantum computers. For example, simulations of Loop Quantum Gravity-derived spin-network states have been carried out \cite{li2019quantum}, and the Sachdev-Ye-Kitaev model \cite{sachdev1993gapless, kitaev2015simple} conjectured to possess a dual gravitational description has been implemented on Google's Sycamore quantum processor \cite{jafferis2022traversable}. Although care should be taken in the interpretation of these experiments \cite{kobrin2023comment}, the promise of implementing toy models of real-world physics on present-day quantum computers deserves further investigation.

Given that the above model starts out from the correspondence between the symmetries of few-qubit systems and relativistic spacetimes, and from there, recovers the standard model gauge symmetries via a `dimensional reduction' that corresponds to fixing a given tensor-product structure, it seems natural to utilize it as a means of simulating these systems on present-day quantum computing hardware. Here, we will only give a brief outline the application of this correspondence to the narrow task of simulating gauge fields.

As noted by von Weizsäcker et al. \cite{weizsacker1958komplementaritat} (see also \cite{lyre2004quantentheorie}, chap. 2.4), we can construct a tensor $F_{\mu\nu}$ from the components of the spinor 
\begin{equation*}
    \ket{\psi}=\begin{pmatrix}
        a \\ b
    \end{pmatrix}
\end{equation*}
of the form:
\begin{equation}\label{eq:emtensor}
    F_{\mu\nu}= \begin{pmatrix}0 & -\frac{1}{2}(a^2-b^2) & -\frac{i}{2}(a^2+b^2) & ab\\ \frac{1}{2}(a^2-b^2) & 0 & iab & -\frac{1}{2}(a^2+b^2)\\ \frac{i}{2}(a^2+b^2) & -iab & 0 & -\frac{i}{2}(a^2 - b^2)\\ -ab & \frac{1}{2}(a^2+b^2) & \frac{i}{2}(a^2-b^2) &0 \end{pmatrix}
\end{equation}

Due to the skew-symmetry $F_{ik}=-F_{ki}$ and self-duality $F_{k0}=if_{lm}$ for $k,l,m$ cyclic, only the three components $F_{0i}$ of this tensor are independent. With $k_\mu$ from Eq.~\eqref{eq:Weyl}, the following hold as algebraic identities:
\begin{equation}\label{eq:Maxwell}
\begin{matrix}
    k_\lambda F_{\mu\nu} + k_\nu F_{\lambda\mu} + k_\mu F_{\nu\lambda} = 0\\
    k^\nu F_{\mu\nu} = 0,
\end{matrix}
\end{equation}
which are formally the (source-free) Maxwell-equations in the momentum-space representation. Consequently, implementing a simulation of Eq.~\eqref{eq:Weyl} on a quantum computer may be used to simulate configurations of the electromagnetic field in free space.

Then, with the correspondence developed in the proceeding, there seems potential to extend this further. A natural possibility here is to use the connection between $5+1$-dimensional massless degrees of freedom and the massive Dirac equation in $3+1$ dimensions (Eq.~\eqref{eq:Dirac}) to enable fermionic simulations.

\section{Conclusion} \label{sec:conc}

The present study draws inspiration and arguments from several disparate sources, aiming to integrate them into a coherent whole. These avenues are, as delineated above:
\begin{itemize}
    \item[$\circ$] The `quantum first' program of deriving physics from scratch, starting from a quantum state in Hilbert space,
    \item[$\circ$] the Kaluza-Klein-like emergence of forces and matter from dimensional reduction,
    \item[$\circ$] the connection between spacetimes of dimensions $2+1$, $3+1$, $5+1$ and $9+1$ and the division algebras $\RN$, $\CN$, $\QN$ and $\ON$,
    \item[$\circ$] the division-algebraic approach to particle physics, and
    \item[$\circ$] the application of division algebras to entanglement theory.
\end{itemize}

In summary, the following picture suggests itself. The background of spacetime, together with its dynamics, emerges from large-scale properties of the area-law contribution to the total entanglement entropy. Area-law violating contributions, which can be viewed as due to an `encoded' state in the sense of quantum error correction, on the other hand, yield gauge- and matter-fields. These can be viewed as what one might picturesquely call `bubbles' of higher dimensional spacetime, which, upon reduction to the $3+1$-dimensional background, yield the familiar gauge symmetries of the Standard Model, along with appropriately transforming matter fields, in a novel realization of the Kaluza-Klein mechanism.

In this vein, we have reviewed how a succession of tensor product spaces of elementary systems gives rise to a succession of constructions based on the Hopf fibration, terminating with the final octonionic case. Using only the postulate of singling out a preferred imaginary direction, motivated by the requirement of representing all higher-order systems within the $3+1$-dimensional spacetime associated to the elementary case of the single qubit, we have seen that the largest of these cases, as shown in $\cite{krasnov2020so9,dubois2016exceptional,todorov2018deducing,todorov2019exceptional}$, gives rise to the exact gauge group of the Standard Model, complete with the correct transformation laws for lepton and quark states (of a single handedness).

The construction can be interpreted in two different ways. In Secs.~\ref{sec:compl}, \ref{sec:quat} and \ref{sec:oct}, the internal symmetries of a left-handed half-generation of fermions were constructed, starting from the equivariance group of the Hopf map. Sec.~\ref{sec:mass} then introduces a way to think about the reduction as yielding the spacetime degrees of freedom of a Dirac fermion, together with the gauge symmetry of the right-handed half.

Consequently, by including one last doubling, taking us to four qubits in the Hilbert space $\CN^{16}\simeq\ON^4$, we sketched a possibility to include both left- and right-handed fermion states, as well as extending the description to massive particles, giving an intrinsic explanation of the chirality of the weak interaction as due to the fact that one set of directions for one of the two Hopf fibrations in this case is taken up as spanning the celestial sphere. 

Thus, a state of four qubits appears to be a microcosm containing many of the ingredients necessary to build up our universe: we get the spacetime and internal symmetries of the right-handed half of one generation of massive Standard Model fermions in $3+1$-dimensions, and the internal symmetries of the left-handed half. Apart from the possibility of model-building, the correspondence described here thus also yields a natural way of implementing physically relevant dynamics with state-of-the-art quantum computers.

Furthermore, as this construction reaches its natural end-point with the octonionic Hopf fibration, the discovery of additional particles or `fifth' forces would not be easily accomodated, and thus, constitute strong evidence against it.

The above does not pretend to give a full theory. Its intent is merely to point out that many of the ingredients making up the most fundamental theory of particle physics, the Standard Model, seem to crop up in an---at first sight---entirely unrelated setting, namely, the structure of entanglement in qubit Hilbert spaces. The characteristics of the Standard Model are often regarded as arbitrary, unsightly, and odd---their independent appearance in two unrelated settings, then, would seem to be doubly odd. Hence, one might venture the more parsimonious hypothesis that there is, in fact, a deep relation between the two areas, whose outline the present account has just begun to trace.

On the basis that this is indeed more than accidental, we have provided a sketch pertaining to how a connection between these fields might be realized, by incorporating ideas from the `quantum first' program. Thus, the present proposal is best understood as attempting to provide a suitable physical background within which the seemingly disconnected strands outlined above fall naturally into place.

The construction as outlined so far, yielding the symmetries of $3+1$-dimensional Minkowski-spacetime and internal symmetries from the symmetries of the state space of three qubits upon a `dimensional reduction' that can be interpreted as singling out a complex direction or a set of degrees of freedom as belonging to the `base' qubit, is mathematically exact. Of course, this has no bearing on whether it is realized in nature; many ideas of great mathematical beauty have failed in the confrontation with experimental data. Thus, the physical relevance of these ideas, if any, remains open to investigation.

Furthermore, there remains much work to be done in putting this idea on the solid footing it needs to be fully convincing. For one, there still seems to be an element of choice in the question of which complex structure to preserve; several inequivalent options are discussed in Ref.~\cite{krasnov2020so9}. There are, of course, also many more elements of the Standard Model our present discussion has not touched upon---such as the values of elementary masses/Yukawa couplings and interaction strengths, the mixing angles, and others. 

One intriguing problem is the inclusion of a Higgs mechanism. Here, we offer up the following observation: the above discussion implicitly assumed that it is possible to associate a single distinguished imaginary axis with each point of the base manifold of the Hopf fibration. However, the nontriviality of the bundle means that this ultimately cannot be done. This is well known in the standard qubit case, where it represents just the impossibility of assigning a consistent phase to every point on the Bloch sphere. 

A difficulty of this sort was already encountered by Finkelstein, Jauch, Schiminovich and Speiser in their original formulation of quaternionic quantum mechanics \cite{finkelstein1963principle}. There, it was proposed that the `special' quaternion imaginary direction should be promoted to a dynamical variable, thus introducing an additional scalar degree of freedom into the theory (a move that was generalized to the octonionic case by Casalbouni, Domokos, and K\"ovesi-Domokos \cite{casalbuoni1976algebraic}). Furthermore, it was shown that, due to this addition, the gauge bosons of the resulting $SU(2)$ gauge theory acquire mass, meaning that the quaternionic field assumes a Higgs boson-like role. It would be highly interesting to investigate this possibility for the present model.

Another salient question lies in the origin of the tensor-product structure we have implicitly assumed in resolving an abstract quantum state into distinct qubits. In ordinary quantum mechanics, systems are largely individuated by their spatial relations---and thus, we assign a tensor product structure according to a split of a system in degrees of freedom \emph{over here} versus \emph{over there}. But in the picture outlined here, we do not have recourse to any a priori spatial structure. What then determines the tensor product structure of the Hilbert space? This is the \emph{quantum factorization problem} \cite{tegmark2015consciousness}, or the question of \emph{quantum mereology} \cite{carroll2021quantum}, and, unless one is willing to settle for considering the tensor product structure as an a priori input into the theory, will need a convincing answer to make the notions outlined so far fully well-defined. Indeed, without an answer to this question, there is in general no fact of the matter regarding whether a system is entangled. 

Recently, Stoica has argued that there is no unique way to reconstruct $3D$ space from the quantum state \cite{stoica20213dspace}. If that is the case, then the present proposal will need to appeal to additional structure. One possibility might be to interpret the proposal in an epistemic way, with a preferred basis being picked out by the actually performed observations. This would put the observer into a central role, opening up a `first-person' account of physics, similarly to the proposal of `law without law' due to M\"uller \cite{mueller2020law}. On this view, the basis, and with it the tensor product structure, then would be a necessary input to the model in order to yield a unique phenomenology.

There are many other intriguing possibilities for future development. For instance, while more complex states do not add more chapters to the story told above, that does not necessarily imply that they can contain nothing but additional copies of the above construction. Further degrees of freedom might exist that, while not having a neat decomposition in terms of ordinary particles as per the present construction, nevertheless add to the total mass-energy content of the universe. As the Standard Model particles can only account for $\sim 15\%$ of this mass-energy content, one might then conjecture that these additional degrees of freedom have something to do with dark matter.

\acknowledgments{The author wishes to thank Kirill Krasnov for helpfull comments to an earlier version of this article, and Tejinder P. Singh for the opportunity to present these results at the OSMU seminar series.}

\bibliography{SMQubit.bib}

\end{document}